# High-mobility field-effect transistor using 2-dimensional electron gas at the LaScO$_3$/BaSnO$_3$ interface

*Hyeongmin Cho, Dowon Song, Youjung Kim, Bongju Kim, and Kookrin Char$^†$*

Institute of Applied Physics, Department of Physics and Astronomy, Seoul National University, Seoul 08826, Korea

ABSTRACT: A novel 2-dimensional electron gas (2DEG) system with high-mobility was discovered at the interface of two perovskite oxides, a polar orthorhombic perovskite LaScO$_3$ (LSO) and a nonpolar cubic perovskite BaSnO$_3$ (BSO). Upon depositing the LSO film on the BSO film, the conductance enhancement and the resulting 2DEG density (n$_{2D}$) was measured. Comparing the results with the previously reported LaInO$_3$/BaSnO$_3$ (LIO/BSO) polar interface, we applied the "interface polarization" model to the LSO/BSO system, in which the polarization exists only over 4 pseudocubic unit cells in LSO from the interface and vanishes afterward like the LIO/BSO interface. Based on the calculations of the self-consistent Poisson-Schrödinger equations, the LSO thickness dependence of n$_{2D}$ of LSO/BSO heterointerface is consistent with this model.

Furthermore, a single subband in the quantum well is predicted. Using the conductive interface and the LSO as a gate dielectric, a 2DEG transistor composed of only perovskite oxides with high field-effect mobility ($\mu_{FE}$) close to 100 cm$^2$·V$^{-1}$·s$^{-1}$ is demonstrated.

# 1. INTRODUCTION

Over a couple of decades, the study of 2-dimensional electron gas (2DEG) emerging at the perovskite oxides interface has attracted much interest, starting with polar LaAlO$_3$/non-polar SrTiO$_3$ (LAO/STO) interface [1]. This heterointerface possesses several fascinating physical phenomena, such as superconductivity, ferromagnetism [2], and Rashba spin-orbit coupling [3-5], etc. These discoveries have sparked widespread interest in SrTiO$_3$-based LaBO$_3$/SrTiO$_3$ polar interfaces with different 3d cations at the B-site, however none of them have broken the limitation of SrTiO$_3$-based interfaces, in particular the instability of its oxygen stoichiometry. Recently, it has been reported that 2DEG is created in another polar/non-polar perovskite interface, LaInO$_3$/BaSnO$_3$ (LIO/BSO) [6-10], which has a very important advantage of being stable even at high temperatures. BSO is a cubic perovskite, with a bulk lattice constant of 4.116 Å and a band gap of 3.1 eV [11]. It can be easily doped with n-type dopants [12,13], and possesses high carrier density ($n_{3D} \sim 10^{20}$ cm$^{-3}$) and high electron mobility ($\mu \sim 320$ cm$^2 \cdot$V$^{-1} \cdot$s$^{-1}$) as well as high oxygen stability [14-17]. The high mobility of BSO at room temperature comes from its small effective mass [18] and the single non-degenerate conduction band of 5s orbitals of Sn [19], whereas many perovskite oxides have triply degenerate bands of d orbitals, resulting in low mobility [20]. LIO is an orthorhombic perovskite of a GdFeO$_3$-type [21] with a band gap of 5.0 eV [22]. Its pseudocubic lattice constant ($a_{pc}$) is calculated as 4.117 Å, which almost matches with the lattice constant of BSO [23,24], 4.116 Å. When depositing an LIO layer on a BSO layer on both STO and MgO substrates, conductance enhancement more than 10$^4$ times were observed at room temperature. Using such LIO/BSO heterostructures, high-mobility ($\mu_{FE} \sim 60$ cm$^2 \cdot$V$^{-1} \cdot$s$^{-1}$) field-effect transistors with a large on/off current ratio ($I_{on}/I_{off} \sim 10^9$) were demonstrated [6,7].

Several experiments have shown that the origin of conductance enhancement at the LIO/BSO interface is neither from formation of oxygen vacancies nor from La inter-diffusion. In addition, no conductance enhancement was observed at the $BaHfO_3/BaSnO_3$ (BHO/BSO) and $SrZrO_3/BaSnO_3$ (SZO/BSO) interfaces, in which BHO and SZO are non-polar perovskite with band gaps of 5.8 eV and 5.6 eV, respectively [6]. Considering 2DEG is generated at the interface with LIO which has a smaller band gap and conduction band offset than those of BHO and SZO, the polar nature of LIO and the resulting polar interface are critical factors in forming a conductive interface. In order to account for the conductance enhancement at the LIO/BSO interface, we introduced the "interface polarization" model where the 2DEG-inducing polarization in the polar orthorhombic LIO exists only near the interface with cubic BSO [8]. Applying this model, we described the variation of the sheet carrier density ($n_{2D}$); the $n_{2D}$ increases for the first 4 unit cells of LIO, reversely decreases for a few nm of LIO, and then approaches a constant value for thicker LIO. Furthemore, the lack of the critical thickness of LIO in forming 2DEG can also be explained by this model that polarization is created from the first unit cell of LIO. However, growth of a polar perovskite on BSO is not a sufficient condition for 2DEG formation. At the $LaGaO_3/BaSnO_3$ (LGO/BSO) interface, another polar/non-polar perovskite interface, it was experimentally confirmed that there was no conductance enhancement although LGO is expected to have a larger band gap and a larger conduction band offset than LIO. LGO is also a polar orthorhombic perovskite but $a_{pc}$ of LGO is calculated as 3.890 Å, much smaller than that of LIO [21,23,25,26]. According to a recently reported study, when a polar orthorhombic perovskite with a lattice constant similar to BSO grows coherently on the BSO, the in-plane lattice constant of the orthorhombic perovskite is pinned with that of the BSO due to coherent epitaxial strain, which induces breaking of inversion symmetry near the orthorhombic/cubic interface, allowing the polar perovskite to possess polarization near the interface with the broken symmetry.  In contrast, when a polar orthorhombic perovskite with large lattice mismatch with BSO grows on the BSO, the in-plane lattice constant of the orthorhombic perovskite cannot be pinned with that of the BSO, resulting in becoming structurally relaxed by forming new dislocations right from the interface.  These dislocations interfere with coherent epitaxial strain near the orthorhombic/cubic interface, which in turn greatly reduces the interfacial polarization at the polar interface.  LIO with $a_{pc}$ = 4.117 Å corresponds to the former case, while LGO with $a_{pc}$ = 3.890 Å corresponds to the latter case [27].

In order to verify the validity of the role the coherent epitaxial growth plays for the interface polarization and the resulting formation of a 2DEG at the polar interface, the studies of the new heterostructures between other polar orthorhombic perovskites and BSO are needed. Herein, we chose to create novel polar/non-polar perovskite heterostructures using $LaScO_3$ (LSO).  In general, rare-earth scandates ($ReScO_3$, with Re = La, Y, Pr, Nd, Gd, Dy, Lu, etc.) are known to be chemically stable at high temperatures and have a high dielectric constant ($\kappa$), and thus have a large potential to be used in field-effect transistor applications [28].  LSO is a polar orthorhombic perovskite [21] with a band gap of 5.8 eV [29-32] and $a_{pc}$ of 4.053 Å [28,33,34], and is also known as a high-$\kappa$ dielectric [28,31,35].  A theoretical prediction that 2DEG formation is possible at the $LaScO_3/BaSnO_3$ (LSO/BSO) interface [36] as well as an experimental result that 2DEG is formed at the $LaScO_3/SrTiO_3$ (LSO/STO) interface [37] have been reported in the past.

In this work, after confirming the property of LSO as an excellent dielectric with high-$\kappa$, electrical properties, we investigated formation of 2DEG at the LSO/BSO interface at room temperature.  We will adopt the "interface polarization" model and apply it to the experimental data of $n_{2D}$ occurring at the LSO/BSO interface.  Furthermore, the energy band structure and the resulting quantum well at the interface will be analyzed through simulation using the self-consistent Poisson-Schrödinger (P-S) formalism. Coherent epitaxial growth of the LSO on the BSO is demonstrated through scanning transmission electron microscopy (STEM) and reciprocal space mapping (RSM), and compared with the case of the LIO/BSO heterostructures. Finally, we fabricate the field-effect transistor (FET) whose entire structure is composed of perovskite oxides using the conductive 2DEC formed in the BSO channel and the LSO as the high-$\kappa$ gate dielectric, and show high field-effect mobility ($\mu_{FE}$) around 100 cm$^2$·V$^{-1}$·s$^{-1}$ at room temperature.

## 2. RESULTS AND DISCUSSIONS

### 2.1. Dielectric properties of LSO film

In order to investigate the dielectric properties of LSO, we built mesa-like capacitors with a 165 nm LSO layer sandwiched between 4% La-doped BSO (BLSO) electrodes with an area of 5.31 × 10$^{-4}$ cm$^2$. Since 4% BLSO is degenerately doped, it can serve as a good metallic contact layer [11,38]. The structure of the sample is detailed in Figure S1. Figure 1a shows the results of the $C_p$ and tan $\delta$ when an AC voltage of the root-mean-square 30 mV with a frequency variation from 10$^3$ to 10$^5$ Hz was applied. $C_p$ remains almost constant over the given frequency range, and tan $\delta$ (composed of all the resistive sources in our 2-point measurement) remains below 0.2, excluding the high-frequency region. From the obtained $C_p$, the dielectric constant ($\kappa$) of the LSO is calculated to be 28, which is almost similar with those of the previously reported LSO films grown on a SiO$_2$ ($\kappa$ ~ 22) [35], a SrRuO$_3$/SrTiO$_3$ substrate ($\kappa$ ~ 26) [31] and a LaAlO$_3$ substrate ($\kappa$ ~ 32) [28], respectively. Next, a DC voltage was applied to measure the leakage current in the capacitor to estimate the breakdown field. From Figure 1b, it can be seen that a sudden and irreversible increase of the current density (J) took place at 3.29 MV cm$^{-1}$, the breakdown field ($E_{BD}$). Although $E_{BD}$ usually improves as the growth is further optimized, from the initial $\kappa$ and $E_{BD}$ measurement of the LSO dielectric film, we can deduce the maximum modulated carrier density $n_{max} = 5.2 \times 10^{13}$ cm$^{-2}$ from the definition $n_{max} = \kappa\varepsilon_0 E_{BD}/e$, where $\varepsilon_0$ and e are the permittivity of the vacuum and the elementary charge, respectively.

To further estimate band alignment between LSO and BSO systems, we analyzed the leakage current density before the dielectric breakdown using the Fowler-Nordheim (FN) tunneling process

$$J \propto E^2 exp\left(\frac{-4\sqrt{2m^*_{LSO}}\Phi^{\frac{3}{2}}}{3e\hbar E}\right), \qquad (1)$$

where J, E, $m^*_{LSO}$, and $\Phi$ are the current density, the electric field, the effective mass of conduction electrons in LSO the layer, and the barrier height between the LSO layer and 4% BLSO electrode, respectively [39]. Here $\Phi$ is derived from the following relation $\Phi = E_{CB,LSO} - E_{F,BLSO}$, where $E_{CB,LSO}$ and $E_{F,BLSO}$ is the energy level of the conduction band minimum (CBM) of LSO and the Fermi level of 4% BLSO, respectively. To calculate $\Phi$ from the above equation, we plotted the ln(J/E$^2$) versus E$^{-1}$ graph, shown in the inset of Figure 1b. The barrier height can be calculated using the linear fit slope of the graph in the high electric field region ranging from 2 MV·cm$^{-1}$ to 3 MV·cm$^{-1}$ and the effective mass of LSO, resulting in $\Phi$ = 0.65 eV. Also, the energy level of the CBM of BSO ($E_{CB,BSO}$) must be calculated to determine the band alignment between LSO and BSO systems. $E_{F,BLSO} - E_{CB,BSO}$ was previously reported to be 0.55 eV [22,40]. Using this result and $\Phi$ obtained from above, the conduction band offset ($\Delta E_{CB}$) between $E_{CB,LSO}$ and $E_{CB,BSO}$ is finally calculated to be about 1.2 eV, which is probably an underestimate since other transport processes than the F-N tunneling often exist before the breakdown. The determined LSO and BSO band alignment can be seen in Figure 1c. Subsequently the valence band offset ($\Delta E_{VB}$) is calculated to be 1.5 eV from the LSO and BSO band gaps. We will show that this large $\Delta E_{CB}$ plays a critical role in forming 2DEG at the LSO/BSO interface.

### 2.2. Electrical transport properties of heterostructures

To measure the electrical properties of LSO/BSO interface such as sheet conductance ($\sigma_s$), sheet carrier density ($n_{2D}$), and electron mobility ($\mu$), we fabricated LSO/BSO heterostructures samples with a similar structure as the previously reported LIO/BSO samples [6]. The 3-dimensional schematic and the top view of the completed sample observed by an optical microscope are shown in Figure 2a and Figure 2b, respectively. Figure 2c shows $\sigma_s$ variation before and after the LSO deposition as a function of the doping rate of the BLSO channel layer. The crimson data points are the experimental results of LSO/BLSO samples, shown against the background of sky blue data points for the results of LIO/BLSO from our previous reports [6]. Looking at the trend of the graph, the conductance enhancement of the LSO/BLSO interface is almost the same as that of the LIO/BLSO interface. Prior to LSO deposition, the BLSO layer by itself remains insulating (sheet resistance > $10^{10}$ Ω) until La concentration of 0.3%. This is caused by the large density of the deep acceptors ($N_{DA, STO} = 6 \times 10^{19}$ cm$^{-3}$) in the BLSO films on STO substrates, including the high density of threading dislocations, that need to be compensated by the La donors in the film. Due to the large deep acceptor density in BSO no conductance appears in the LSO/BSO(undoped) interface, as in the LIO/BSO(undoped) interface. For the LIO/BLSO interface grown on MgO substrate, however, the conductance increases by about a factor of $10^5$ at the LIO/BSO(undoped) interface due to the relatively smaller deep acceptor density ($N_{DA, MgO}$ = 4 × $10^{19}$ cm$^{-3}$) of the BSO films on the MgO substrate [17]. Therefore, if the LSO/BLSO heterostructures are made on the MgO substrate, it can be predicted that there will be an increase of conductance at the LSO/BSO(undoped) interface. To rule out the possibility of conduction enhancement from oxygen vacancies of the BLSO layer which can be generated during the deposition process, an LSO/BLSO(0.3%) sample was annealed in the O$_2$ environment at 400 °C for 5 hours. Referring to Figure S2, the increased sheet conductance by forming an interface with 100 nm LSO hardly changed after annealing in O$_2$ environment, meaning that the effects of the oxygen vacancies on the LSO/BLSO interface are almost negligible, unlike the LAO/STO interface where oxygen annealing changes its properties [41,42]. The following Figure 2d shows $n_{2D}$ and µ generated at the LSO/BLSO interface, which are also compared to those generated at the LIO/BLSO interface. Here we can see that µ increases with increasing $n_{2D}$, consistent with the previously reported carrier behavior in BLSO films, in which µ is limited by charged impurity scattering, such as threading dislocations [11,14,16,38]. The difference in $n_{2D}$ between the two types of interfaces is very small, but µ at the LSO/BLSO interface seems a little smaller. It will be discussed in the structural analysis section to see if the relatively lower mobility of the LSO/BLSO interface is related with the strain difference between the interfaces of the LSO/BSO and LIO/BSO.

Next, the changes of $\sigma_s$, $n_{2D}$, and µ were measured while laying down LSO layer unit cell by unit cell on 0.3% BLSO layer, and the results are shown in Figure 3a. The thickness dependence of the LSO/BLSO(0.3%) interface also shows a similar thickness variation as that of the LIO/BLSO(0.3%) interface. Even if only 1 unit cell of the LSO is deposited, a rapid increase of $\sigma_s$ can be confirmed. $\sigma_s$ and $n_{2D}$ continue to increase until 4 unit cells of LSO are stacked, and then begin to gradually decrease as 5 more unit cells are stacked. $n_{2D}$ remains almost unchanged from 6 more unit cells until hundreds of unit cells are stacked.

In order to understand LSO/BLSO interfaces more quantitatively, we have applied the previously reported "interface polarization" model, which successfully described the conductance behavior for the LIO/BLSO(0.3%) interface [8]. For this, the self-consistent Poisson-Schrödinger (P-S) band calculator developed by G. Snider was used [43]. From such P-S simulation, not only the band bending and the charge distribution at the given heterointerfaces can be obtained, but also the subband characteristics can be calculated. The important input material parameters for this simulation are energy gap ($E_g$), conduction band offset to BSO ($\Delta E_{CB}$), dielectric constant ($\kappa$), and effective mass ($m_e^*$), etc. These values are summarized in Table 1 [8,11,22,31,44,45], and the heterostructures used in the simulation are shown in Figure S3.

| Material | $E_g$ (eV) | $\Delta E_{CB}$ to BSO (eV) | $\kappa$ | $m_e^*/m_e$ | $E_D$ (eV) | $E_A$ (eV) |
|---|---|---|---|---|---|---|
| BSO | 3.1 | – | 20 | 0.42 | -0.63 | 1.55 |
| LIO | 5.0 | 1.6 | 38 | 0.46 | 2.5 | – |
| LSO | 5.8 | 1.2 | 28 | 0.3 ~ 0.4 | 2.5 | – |

**Table 1.** Material parameters used for BSO, LIO, and LSO.

Based on the experimental results in Figure 3a, where the $n_{2D}$ increases while stacking 4 LSO unit cells, we assumed an interface polarization in LSO which exists only in 4 unit cells near the interface and disappears after that [8]. In the case of the LIO/BLSO(0.3%) interface which also showed the maximum $n_{2D}$ at 4 unit cell thickness of LIO, interface polarization values of 65/65/25/10 $\mu C/cm^2$ over the LIO 4 unit cells were found to fit with the experimental $n_{2D}$. Taking account of this, we performed several P-S simulations for the best fit with the experimental $n_{2D}$ in Figure 3a and reached the interface polarization values of 61.5/61.5/23/10 $\mu C/cm^2$ in the LSO, as shown in Figure 3b and Figure 3c. By introducing large polarization values from the first unit cell of the LSO, the absence of a critical thickness of the LSO in forming 2DEG can be explainable. Subsequently, the decreasing $n_{2D}$ beyond 4 unit cells of LSO is due to the reversing potential starting at the end of the interface polarization. In order to screen such potential, positive charges will be needed, hence the introduction of the deep donors. These deep donors can provide the positive charges for screening when they are excited when the Fermi level ($E_F$) is lower than the deep donor state level near the interface. Once the screening is finished by such deep donors, the band bending stops. We found a deep donor density ($N_{DD}$) of $2.5 \times 10^{20}$ $cm^{-3}$ at the level of 2.5 eV below its conduction band minimum in the LSO can fit our experimental $n_{2D}$ very well. We have placed the deep donor level at 2.5 eV below the conduction band minimum just as in the case of LIO [8]. These deep donor states can very well be the oxygen vacancy states in LSO or LIO, which can be ionized to become positively charged when the $E_F$ is lower than the oxygen vacancy state level; the $N_{DD}$ of $2.5 \times 10^{20}$ $cm^{-3}$ in the LSO corresponds to $LaScO_{2.991}$. Judging from the $O_2$ environment annealing results in Figure S2, the deep donor states in LSO seem stable and intrinsic, same as in the LIO/BLSO cases.

The calculated band diagrams of the LSO/BLSO(0.3%) heterostructures when the LSO thickness is 100 nm, such as the structure in Figure S3, are shown in Figure 4. In this simulation, $N_{DD}$ values in Figure 3c were used. From the figure, it can be seen that the interface polarization, along with the conduction band offset, largely tilts the bands of the LSO at the interface. Although the interface polarization values (61.5/61.5/23/10 µC/cm$^2$) and the conduction band offset to BSO of LSO(1.2 eV) are smaller than those of LIO (65/65/25/10 µC/cm$^2$ and 1.6 eV), the relatively lower $\kappa$ of LSO ($\kappa \sim 28$) than that of LIO ($\kappa \sim 38$) makes band bending on the LSO side easier than for LIO, resulting in the same 2DEG density as in the case of the LIO/BSO interface. Taking a closer look, in the region around 10 to 30 Å, the $E_F$ becomes lower than deep donor states ($E_{DD}$), so that the $E_{DD}$ are ionized. The resulting free electrons released from the ionized $E_{DD}$ move toward the 0.3% BLSO layer, creating a confined 2DEG quantum well with about 2 nm width near the interface. The width of about 2 nm is consistent with the 2D and 3D carrier density dependence of the thermopower measurement of the BSO thin film [18]. This is similar to the mechanism of carrier transfer in conventional modulation-doped semiconductor heterojunctions. For example, in the case of the AlGaAs/GaAs heterostructures, Si is intentionally doped into the AlGaAs layer around 50 to 150 Å away from the interface, where Si donor level is known to be located at 0.1 eV below the conduction band of AlGaAs. When this Si-doped AlGaAs makes heterojunctions with GaAs, Si dopants in the AlGaAs layer are ionized by band bending and the free electrons generated thereby move to the GaAs layer [46,47]. In the same way, it is likely that the deep donors present in the LSO layer play a similar role to the Si dopants present in the AlGaAs layer. Discussions on the similarity of the principle of 2DEG formation in conventional semiconductor heterojunctions (AlGaAs/GaAs, AlGaN/GaN, and MgZnO/ZnO) and perovskite heterostructures are covered in detail in a recent paper [48].

In the inset of Figure 4, n$_{3D}$ in the quantum wells are shown with a single occupied subband ($E_1$). Such single subband occupation differs from the LAO/STO interface with multiple subbands [49]. Calculations using $E_F - E_1$ (~ 0.1 eV), and the 2D density of states, $m_e^*/\pi\hbar^2$, of about $2 \times 10^{14}$ cm$^{-2}$·eV$^{-1}$ in BSO predict that LSO/BSO can accommodate the n$_{2D}$ of about $2 \times 10^{13}$ cm$^{-2}$, which is quite consistent with our experimental values. These non-degenerate quantum states at the LSO/BSO interface can make it easier to observe quantum phenomena such as the Shubnikov-de Haas oscillation [50] or Quantum Hall effect [51] in 2DEG systems, when the low temperature mobility improves by reducing the scattering by defects such as the threading dislocations.

### 2.3. Structural analysis of heterostructures

In order to investigate the structural properties, scanning transmission electron microscopy (STEM) was performed. The LSO/BSO layer including the STO substrate can be seen in Figure 5a. The measured sample was of a structure of 78 nm LSO film deposited on the 17 nm BSO film on the TiO$_2$-terminated STO (001) substrate. It can be confirmed that the large density of the threading dislocations (TDs) created in BSO on STO extend beyond the BSO layer to the LSO layer. It has been reported that the TD density of BSO deposited on the STO substrate is about $5 \times 10^{10}$ cm$^{-2}$ [38,52]. A similar TD density can be seen in Figure 5a. Next, an image of a higher-resolution LSO/BSO interface can be found in Figure 5b. Here too, the TDs created in the BSO layer are maintained up to the LSO layer, but no new formation of TDs was found at the interface, implying that the growth of LSO film is coherently epitaxial and not creating new dislocations near the LSO and BSO interface. This is in contrast to the case of LGO/BSO heterostructures, where new dislocations, not being extended from the BSO layer, start to form right from the LGO/BSO interface, preventing the coherent epitaxial growth [27]. In the case of LIO/BSO, we have already shown that the coherent epitaxial growth between the orthorhombic and cubic structures leads to structural modifications of the polar perovskite near the interface, resulting in interface polarization, which in turn can form a 2DEG state [8,27]. Referring to the measurement results of aberration-corrected transmission electron microscopy [53] and synchrotron scattering [54], this orthorhombic/cubic epitaxial strain appears to exist in a few unit cells along the direction of epitaxial growth near the interface. Similarly, the coherent growth of LSO on BSO seems to contribute to generating the 2DEG state at the interface. Energy dispersive spectroscopic analysis of the LSO/BSO interface can be found in Figure S4 in the supporting information.

Next, Figure 5c shows the reciprocal space mapping (RSM) scan around the peaks of the (103) plane of a LSO(56 nm)/BSO(15 nm) heterostructure on a STO substrate. All intensities of the 3 peaks are well confined in reciprocal space, and there is no polycrystalline growth. Among the two overlapping picks, the upper peak belongs to the pseudocubic (103) plane of LSO and the lower peak belongs to the cubic (103) plane of BSO. The reciprocal space vectors $Q_x = 1.5325$ $Å^{-1}$ and $Q_z = 4.6181$ $Å^{-1}$ correspond to in-plane lattice and out-of-plane lattice parameters $a_{LSO} = 4.1001$ Å, $c_{LSO} = 4.0816$ Å of the pseudocubic (103) plane of the LSO. Subsequently, the reciprocal space vectors $Q_x = 1.5319$ $Å^{-1}$ and $Q_z = 4.5724$ $Å^{-1}$ correspond to in-plane lattice and out-of-plane lattice parameters $a_{BSO} = 4.1014$ Å and $c_{BSO} = 4.1224$ Å of the cubic (103) plane of the BSO. Considering that the bulk lattice constants of LSO and BSO are known as $a_{pc} = 4.053$ Å, $a_c = 4.116$ Å, respectively [11,33,34], the entire LSO is coherently grown on the BSO with its in-plane lattice constant almost pinned with that of BSO, at least near the interface. While the in-plane lattice of the LSO is tensilely strained and coherently pinned with BSO, the out-of-plane lattice constant of the LSO is also expanded from the known bulk value. We have found that as the oxygen partial pressure during deposition increased, the proportion of La in the LSO film increased, leading to an increase in the unit cell volume, the details of which will need more investigation in the future. The reason for the broader pick along the $Q_x$ direction of LSO than that of BSO is that there are 3 differently oriented domains of orthorhombic LSO when grown on cubic BSO, which causes the circumferential spread of LSO in RSM. This broadening can also be identified in the RSM (103) scan result of LIO/BSO heterostructures on the STO substrate in Figure 5d [7,22]. There, among the two overlapping picks, the upper peak belongs to the cubic (103) plane of the BSO and the lower peak belongs to the pseudocubic (103) plane of the LIO. The reciprocal space vectors $Q_x = 1.5241$ Å$^{-1}$ and $Q_z = 4.5441$ Å$^{-1}$ correspond to in-plane lattice and out-of-plane lattice parameters $a_{LIO} = 4.1227$ Å, $c_{LIO} = 4.1481$ Å of the pseudocubic (103) plane of the LIO. The BSO (103) peak is almost overlapped with the LIO pseudocubic (103) peak, so it cannot be accurately distinguished.

However, given the bulk lattice constant of BSO, the in-plane lattice of the LIO appears to be compressively strained from BSO as opposed to that of the LSO. Here, we can consider that the difference in strain direction between the two interfaces (LSO/BSO and LIO/BSO interfaces) makes the mobility difference of the 2DEG occurring at the two interfaces. Considering that the resulting 2DEG state is created in a 2 nm BSO layer near the interface and that recent theoretical studies have reported that the mobility increases when BSO is subjected to tensile strain [55], the BSO under tensile strain from the LIO at the LIO/BSO interface is considered to have relatively higher mobility than BSO under compressive strain from the LSO at the LSO/BSO interface. This agrees with the results of the electrical transport properties in Figure 2 and Figure 3.

### 2.4. Field-effect transistor

A field-effect transistor (FET) was fabricated using 0.2% BLSO as a channel layer and LSO as a gate oxide. The 3-dimensional schematic, the top view of the completed FET observed by an optical microscope, and the structural cross-sectional view are shown in Figure 6a, Figure 6b, and Figure S5, respectively. The channel width (W) is 140 μm, the channel length (L) is 60 μm, and the width of the line-patterned gate electrode is 70 μm, which can sufficiently overlap the channel length. Unlike the above experiments using 0.3% BLSO, 0.2% BLSO was selected as the channel layer in FET to create an accumulation mode FET and reduce the impurity scattering in the 2DEG channel at the interface with LSO. Indeed, after depositing the 275 nm thick LSO layer, the resistance of the channel layer decreased from $1.5437 \times 10^9$ Ω to $6.2786 \times 10^5$ Ω due to the formation of 2DEG at the LSO/BLSO(0.2%) interface. The output characteristics of the FET are shown in Figure 6c. Source-drain voltage ($V_{DS}$) was applied up to 10 V while gate voltage ($V_{GS}$) varied from 25 to -5 V with an interval of 5 V. As the $V_{GS}$ decreases from 25 to -5V, source-drain current ($I_{DS}$) also decreases with $V_{GS}$, consistent with the behavior of a standard n-type FET. Each of the 6 solid lines shows the same characteristics that the $I_{DS}$ is proportional to the low $V_{DS}$ and is pinched off at the drain end to saturate, whereby $V_{DS}$ no longer affects the channel in the high $V_{DS}$ limit. Next, the transfer characteristics of the FET are shown in Figure 6d. Based on the findings above, $V_{GS}$ was swept from 25 to -5 V with $V_{DS} = 1$ V applied, which belongs to the linear region. The figure shows clear switching on/off with little leakage through the gate oxide, with the $I_{on}/I_{off}$ ratio of about $10^6$. The field-effect mobility ($\mu_{FE}$) was calculated using the relation

$$\mu_{FE} = \left(\frac{Lt}{W\kappa\varepsilon_0 V_{DS}}\right)\frac{\partial I_{DS}}{\partial V_{GS}}, \qquad (2)$$

where $L$, $t$, $W$, and $\varepsilon_0$ are the channel length, the thickness of the gate oxide, the channel width, and the permittivity of the vacuum, respectively. The maximum $\mu_{FE}$ was calculated to be about 98.7 cm$^2$·V$^{-1}$·s$^{-1}$, which is the highest value reported to date in FETs based on the BLSO channel layer using various gate oxides such as Al$_2$O$_3$ [56], HfO$_2$ [17,57-60], BaHfO$_3$ [40], and even LaInO$_3$ [7,9,22]. Although the transconductance ($g_m$), defined as $g_m = \partial I_{DS}/\partial V_{GS}$, is slightly higher in the previously reported FETs using the LIO gate oxide than in the LSO gate oxide device, the relatively lower $\kappa$ of the LSO gate oxide than that of LIO renders the FET to show greater µ$_{FE}$. Taken together, such high performance of the device can be attributed to the high conductivity of the channel due to the formation of the 2DEG at the LSO/BLSO(0.2%) interface, as well as the gate dielectric properties of LSO, which is chemically stable at high temperatures with high-$\kappa$, being one of the rare-earth scandates.

## 3. CONCLUSION

In summary, we have shown a new 2DEG system is created at the polar interface between LSO and BSO. The conductance enhancement at the interface was explained by P-S simulations that employed the interface polarization model in which the polarization exists in 4 pseudocubic unit cells of the LSO from the interface and vanishes after that. The calculation also predicts there is a non-degenerate single subband occupation at the LSO/BSO interface, and this feature may be advantageous for observing quantum phenomena such as Quantum Hall effect. STEM and RSM studies of the interface suggest the LSO film growth is coherently epitaxial on the BSO, as is the case with the LIO film growth on the BSO. After investigating the excellent properties of LSO as a high-$\kappa$ dielectric, we have fabricated high-mobility FET using the 2DEG channel at room temperature notwithstanding the high-density threading dislocations in the BSO channel which impede the mobility of the channel. Once the low temperature mobility is improved by reducing the defect scattering, the 2DEG state of the LSO/BSO heterostructures will open up the possibilities to observe new quantum phenomena as well as to fabricate high-power field-effect devices.

## 4. EXPERIMENTAL DETAILS

**4.1. Heterostructure samples fabrication.** In all experiments, samples were grown on $TiO_2$-terminated STO (001) substrates by pulsed laser deposition using a KrF excimer laser ($\lambda \approx 248$ nm; Coherent) with following conditions: the temperature of 750 °C, the oxygen pressure of 0.1 Torr, the target-to-substrate distance in the range of 50 – 55 mm, and the energy fluence in the range of 1.3 – 1.5 J/cm$^2$ during deposition. All targets were provided by Toshima Manufacturing Co. in Japan. **a)** LSO capacitors: a 150 nm 4% BLSO layer as the bottom electrode was grown on the entire area of the 5 mm × 5 mm TiO$_2$-terminated STO (001) substrate. Next, a 165 nm LSO dielectric layer was grown using a rectangular-shaped Si stencil mask. Last, 150 nm 4% BLSO layers as the top electrodes were grown using a stainless steel mask with a total of 21 (7 × 3) circular holes, which builds 21 (7 × 3) mesa-like capacitors on a single chip. **b)** LSO/BSO heterostructures samples (Figure 2 and Figure 3 samples): First, a 150 nm un-doped BSO buffer layer was grown on the entire area of the 5 mm × 5 mm TiO$_2$-terminated STO (001) substrate. Next, a 12 nm BLSO channel layer was deposited using a 3 mm × 3 mm square-patterned Si stencil mask. Subsequently, 60 nm 4% BLSO contact layers were deposited using another type of Si stencil mask on the four corners of the BLSO channel layer. Last, LSO layers were deposited using the same 3 mm × 3 mm mask, which covers the entire BLSO channel layer. This process involves 4 high-temperature cycles and 3 air exposures, during which the surface of the BSO and BLSO layers is terminated with SnO$_2$. **c)** Field-effect transistor: First, a 150 nm un-doped BSO buffer layer was grown on the entire area of the 5 mm × 5 mm TiO$_2$-terminated STO (001) substrate. Next, a 12 nm 0.2% BLSO channel layer was deposited using a Si stencil line mask with a channel width of 140 μm. Subsequently, 60 nm 4% BLSO source-drain contact layers were deposited using a butterfly-shaped stainless steel mask making the channel length 60 μm. After the growth of contact layers, a 275 nm LSO dielectric layer was grown using a rectangular-shaped Si stencil mask. In the final step, a 4% BLSO gate contact layer was grown on the top of the LSO dielectric layer using a Si line mask with a width of 70 μm, which covers the entire channel length with some overlap with the source and drain electrodes.

**4.2. Electrical properties measurement.** In order to measure the capacitance, an AC voltage of the root-mean-square 30 mV with a frequency variation from $10^3$ to $10^5$ Hz was applied to obtain the admittance ($Y$) and the phase shift ($\pi$-$\delta$/2), from which the parallel capacitance ($C_p$) and dissipation factor (tan δ) were derived from the following relation: $|Y| = \omega C_p \sqrt{1 + tan^2 \delta}$. And, to measure electrical transport properties such as sheet conductance ($\sigma_s$), sheet carrier density ($n_{2D}$), and electron mobility (μ), the Van der Pauw method and Hall measurement were conducted at room temperature. All of the above electrical properties were measured using a Keithley 4200 SCS.

**4.3. Microstructural measurement.** In order to investigate structural properties, a spherical aberration-corrected scanning transmission electron microscope (Cs-STEM) was used using a JEM-ARM200F (JEOL Ltd, Japan) equipped with a cold field emission gun (Cold FEG) operating at an electron acceleration voltage of 200 kV. Specimens to be used for STEM were prepared primarily by focused ion beam milling (Helios 650 FIB; FEI, USA) and secondarily thinned by focused Ar-ion beam nano milling (M1040 Nano Mill; Fischione, USA). Next, reciprocal space mapping (RSM) analyses were performed using a SmartLab with a Cu K α-1 source (λ = 1.5406 Å; Rigaku, Japan) at room temperature. An X-ray CBO system, a Ge (220) 2-bounce monochrometer, and a 1-dimensional semiconductor array detector (hybrid photon counting detector; HyPix-3000) were used for the high-resolution crystalline qualities.

**4.4. Self-consistent Poisson-Schrödinger simulation.** The Poisson-Schrödinger simulation (P–S simulation) computes the 1-dimensional Poisson and Schrödinger equations self-consistently in an iterative manner. First, we use a trial potential in the Schrödinger equation to obtain the eigenfunctions and energy eigenvalues. Then, we put these eigenfunctions and energy eigenvalues into the Poisson equation and get a corrected potential as a solution. Last, we take this modified potential back into the Schrödinger equation and repeat this process until the solution again satisfies both equations. From such P-S simulation, the band bending, the charge distribution, and the subband characteristics in the given semiconductor heterostructures can be obtained.

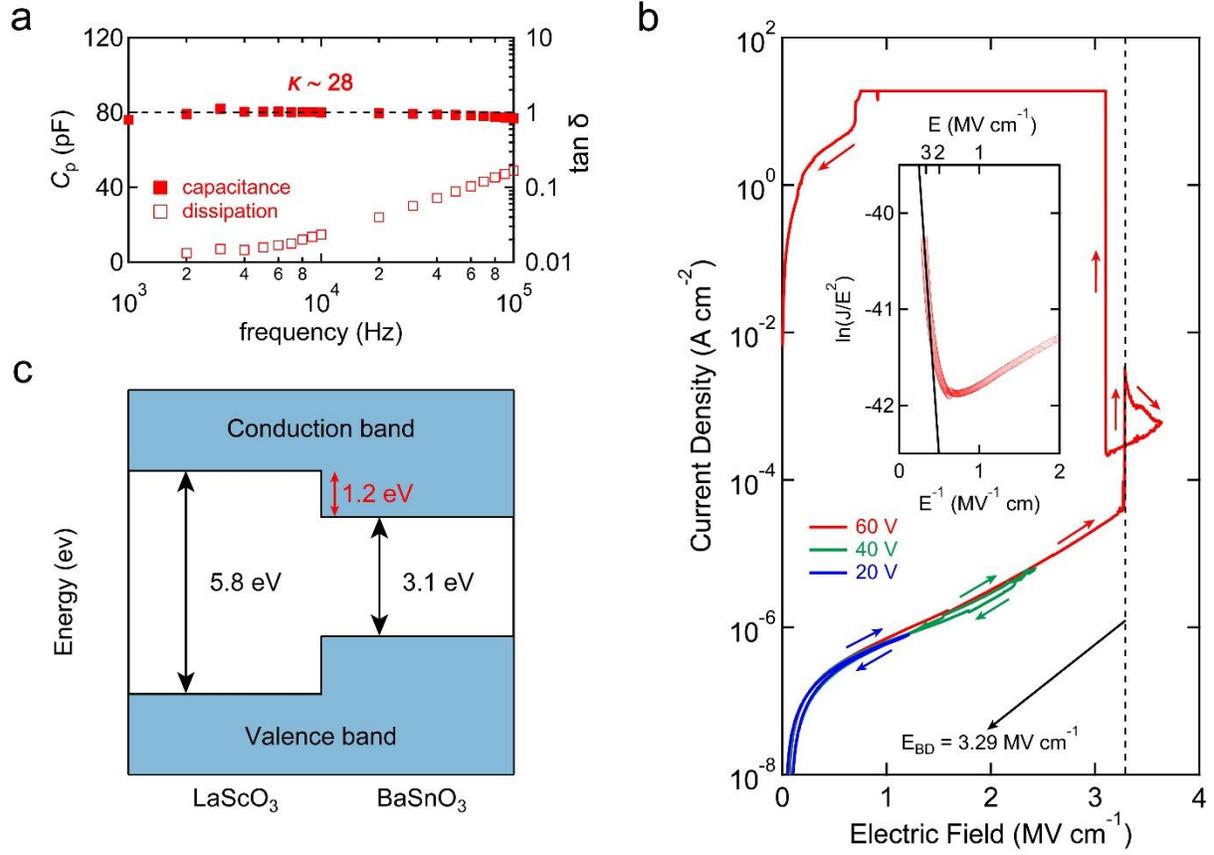

**Figure 1.** The dielectric properties of LSO. (a) The capacitance and dissipation factor of the 165 nm LSO dielectric layer sandwiched between 4 % BLSO electrodes was measured with respect to the applied frequencies of AC voltage. $\kappa$ was calculated from the measured capacitance. (b) The leakage current density (log J) versus electric field (E) characteristics plot of the LSO dielectric layer. $E_{BD}$ is determined from the rapidly increasing J. In the inset the $\ln(J/E^2)$ versus $E^{-1}$ graph is plotted for analysis by the FN tunneling process in the LSO dielectric layer. (c) The diagram of the band alignment between LSO and BSO systems. $\Delta E_{CB}$ is derived from the experimental results in Figure 1a and Figure 1b.

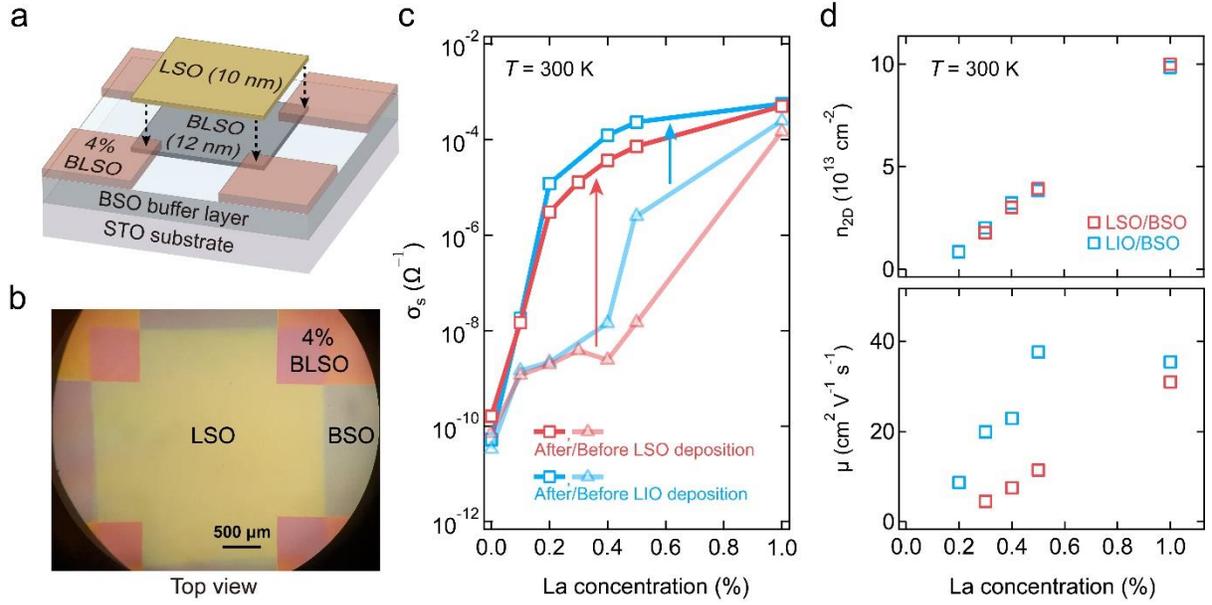

**Figure 2.** (a, b) A 3-dimensional view and a top view of the LSO/BLSO interface. (c) Sheet conductance ($\sigma_s$) before and after LSO deposition as a function of the La concentration of the BLSO layer (crimson color), and comparison with those of LIO deposition (sky blue color). (d) Sheet carrier density ($n_{2D}$) and electron mobility ($\mu$) generated at the LSO/BLSO interface (crimson empty squares) and LIO/BLSO interface (sky blue empty squares).

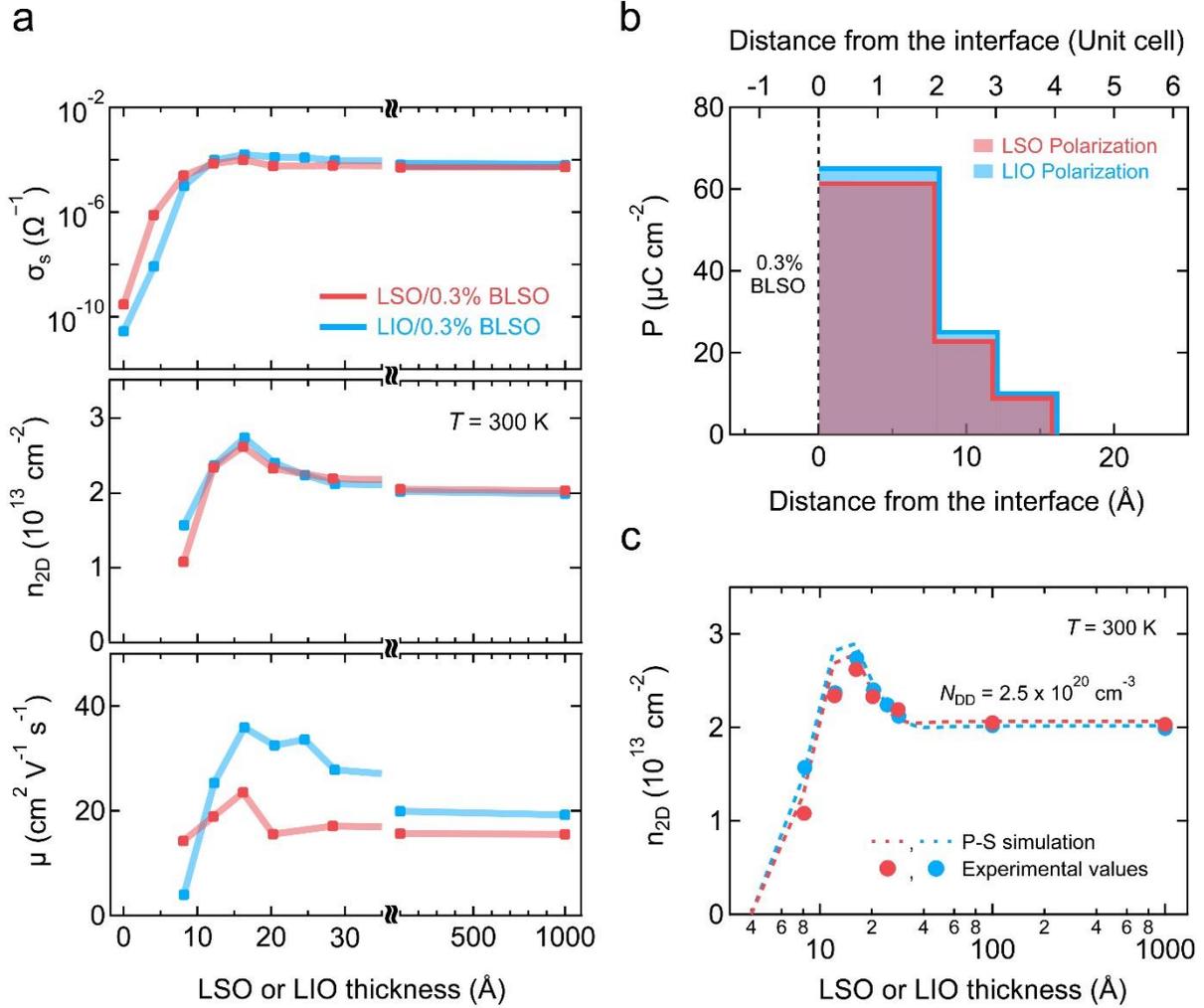

**Figure 3.** (a) Changes of sheet conductance, sheet carrier density, and electron mobility generated at the LSO/BLSO(0.3%) and LIO/BLSO(0.3%) interfaces as a function of the LSO (crimson color) and LIO (sky blue color) thickness. (b) Self-consistent Poisson-Schrödinger simulation for the LSO/BLSO(0.3%) and the LIO/BLSO(0.3%) heterointerfaces. Interface polarization values used for each simulation. The polarization exists over 4 pseudocubic unit cells from the interface and disappears after them. (c) Comparison of $n_{2D}$ by experiments (circles) with those by simulation (dotted lines). The deep donor density ($N_{DD}$) in the LSO is set to $N_{DD} = 2.5 \times 10^{20}$ cm$^{-3}$.

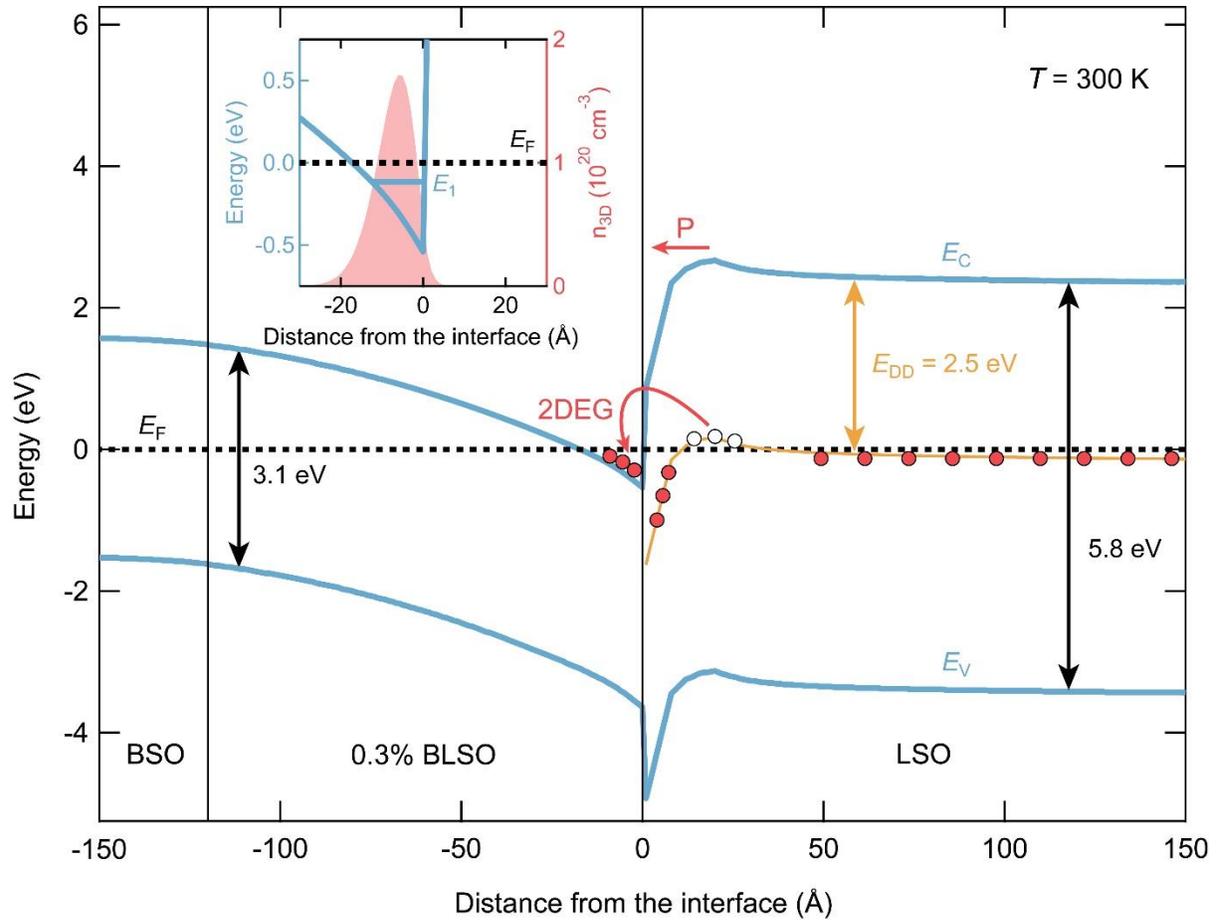

**Figure 4.** A band diagram of the LSO/BLSO(0.3%) calculated by the Self-consistent Poisson-Schrödinger simulation. The sample structure in the simulation is the same as Figure 2a when the LSO thickness is 100 nm, and the $N_{DD}$ value in Figure 3c was applied. In the region around 10 ~ 30 Å, the Fermi level ($E_F$) becomes lower than deep donor states ($E_{DD}$), in which the $E_{DD}$ are ionized. The resulting free electrons released from the ionized $E_{DD}$ move toward the 0.3% BLSO, producing the quantum well. The inset is an enlarged image near the interface with n$_{3D}$ and a single occupied subband ($E_1$).

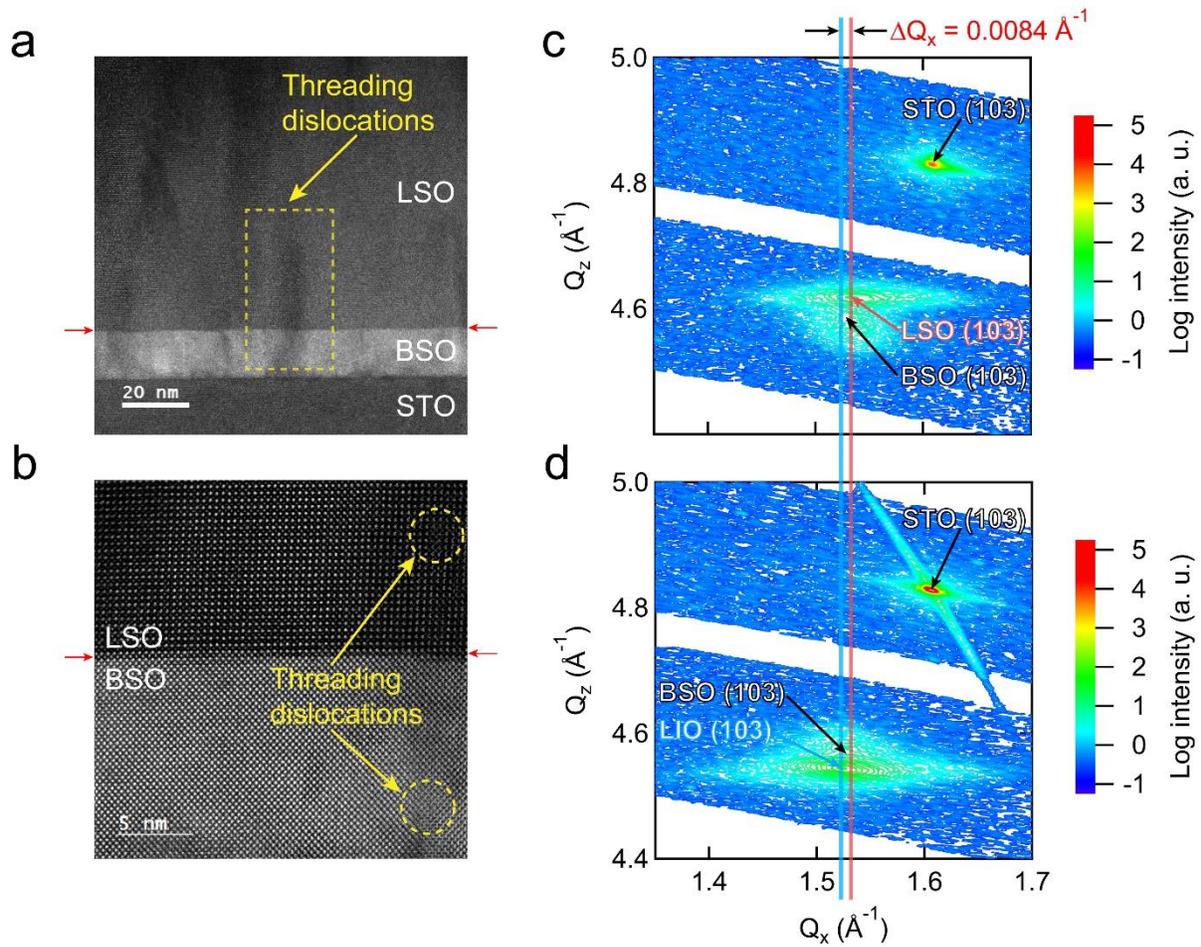

**Figure 5.** (a, b) Cross-sectional high-angle annular dark field scanning transmission electron microscope (HAADF-STEM) images of LSO/BSO films on STO substrate and LSO/BSO interface. In the yellow dashed area, a threading dislocation created in the BSO film continues in the LSO film. But, no new formation of dislocations was found at the interface, indicating the coherent epitaxial growth of LSO on BSO. The Sc atoms look distinct from the Sn atoms unlike the case of the LIO/BSO interface where it is difficult to distinguish between Sn and In. (c, d) The RSM (103) scans results of LSO/BSO films and LIO/BSO films on the STO (001) substrate. The LSO (103) peak is on the crimson vertical line, and the LIO (103) peak is on the sky blue vertical line. The BSO (103) peaks in both figures are located between the two lines.

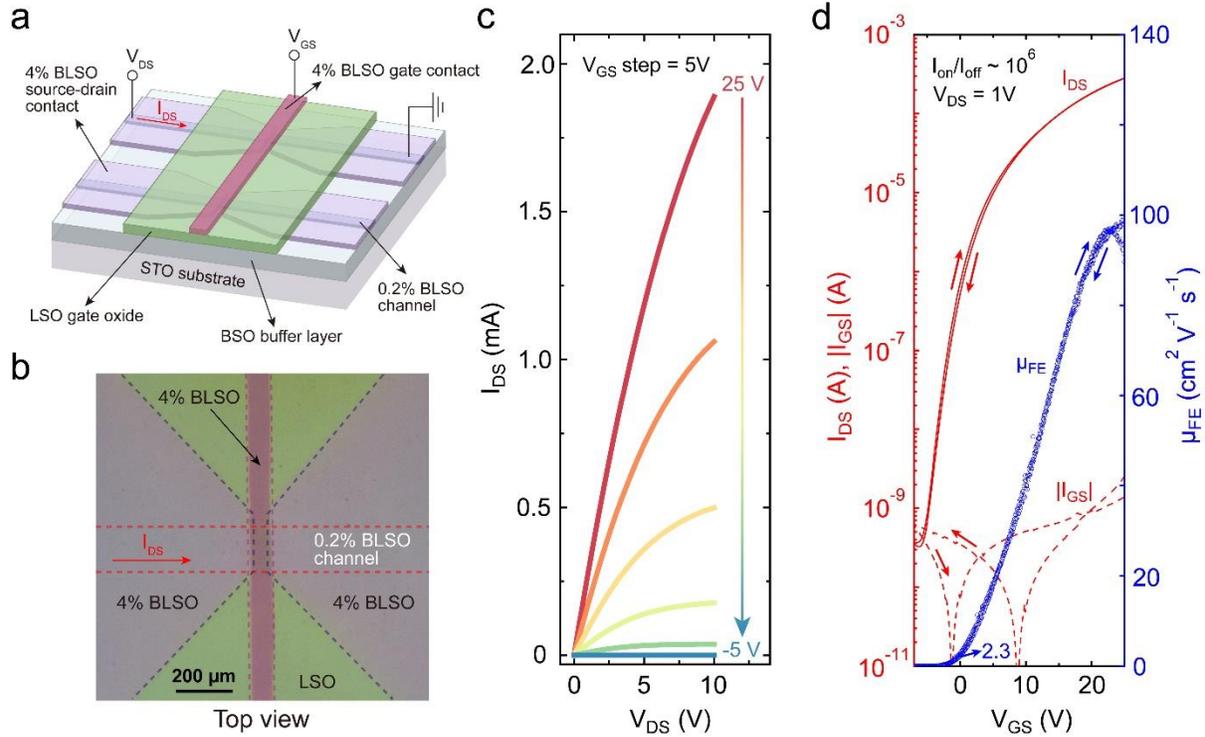

**Figure 6.** A field-effect transistor using 0.2% BLSO as a channel layer and LSO as a gate oxide. After depositing the LSO layer on the 0.2% BLSO layer, there is conductance enhancement from the formation of 2DEG. (a) Schematic of the structure of the device. (b) The top view of the device pictured by an optical microscope. Dashed lines are plotted to distinguish each deposited layer. (c) The output characteristics of the device varying the gate voltage ($V_{GS}$) from 25 to -5 V with a 5 V interval. (d) The transfer characteristics of the device with the source-drain voltage ($V_{DS}$) maintaining 1 V. The source-drain current ($I_{DS}$) is represented by a red line and the leakage current ($I_{GS}$) is represented by a red dashed line. The calculated field-effect mobility ($\mu_{FE}$) is represented by the blue circles. The device shows $I_{on}/I_{off}$ ratio about $10^6$ and the maximum $\mu_{FE}$ close to 100 cm$^2$·V$^{-1}$·s$^{-1}$ at room temperature.

## ASSOCIATED CONTENT

**Supporting Information.** The following files are available free of charge.

A 3-dimensional and a top view of the capacitors, the sheet conductance variations of LSO/BLSO(0.3%) interface after annealing, parameters in the P-S simulation, EDS analysis of the LSO/BSO interface, and the schematics of the field effect transistor.

## AUTHOR INFORMATION

**Corresponding Author**

Kookrin Char – Institute of Applied Physics, Department of Physics and Astronomy, Seoul National University, Seoul 08826, Korea; orcid.org/0000-0001-6060-5448; E-mail: kchar@snu.ac.kr

**Authors**

Hyeongmin Cho – Institute of Applied Physics, Department of Physics and Astronomy, Seoul National University, Seoul 08826, Korea; orcid.org/0000-0002-7256-4812

Dowon Song – Institute of Applied Physics, Department of Physics and Astronomy, Seoul National University, Seoul 08826, Korea; orcid.org/0000-0002-0142-2761

Youjung Kim – Institute of Applied Physics, Department of Physics and Astronomy, Seoul National University, Seoul 08826, Korea; orcid.org/0000-0002-8233-536X

Bongju Kim – Institute of Applied Physics, Department of Physics and Astronomy, Seoul National University, Seoul 08826, Korea; orcid.org/0000-0002-5560-6144

**Notes**

The authors declare no competing financial interest.

ACKNOWLEDGMENTS

Measurements of electrical properties were conducted using a Keithley 4200 SCS at IBS Center for Correlated Electron Systems, Seoul National University.

REFERENCES

(1) Ohtomo, A.; Hwang, H. Y. A high-mobility electron gas at the LaAlO$_3$/SrTiO$_3$ heterointerface. *Natrue* **2004,** *427*, 423-426.

(2) Bert, J. A.; Kalisky, B.; Bell, C.; Kim, M.; Hikita, Y.; Hwang, H. Y.; Moler, K. A. Direct imaging of the coexistence of ferromagnetism and superconductivity at the LaAlO$_3$/SrTiO$_3$ interface. *Nat. Phys.* **2011,** *7*, 767-771.

(3) Caviglia, A. D.; Gabay, M.; Gariglio, S.; Reyren, N.; Cancellieri, C.; Triscone, J. M. Tunable Rashba spin-orbit interaction at oxide interfaces. *Phys. Rev. Lett.* **2010,** *104*, 126803.

(4) Lesne, E.; Fu, Y.; Oyarzun, S.; Rojas-Sanchez, J. C.; Vaz, D. C.; Naganuma, H.; Sicoli, G.; Attane, J. P.; Jamet, M.; Jacquet, E.; George, J. M.; Barthelemy, A.; Jaffres, H.; Fert, A.; Bibes, M.; Vila, L. Highly efficient and tunable spin-to-charge conversion through Rashba coupling at oxide interfaces. *Nat. Mater.* **2016,** *15*, 1261-1266.

(5) Song, Q.; Zhang, H.; Su, T.; Yuan, W.; Chen, Y.; Xing, W.; Shi, J.; Sun, J.; Han, W. Observation of inverse Edelstein effect in Rashba-split 2DEG between SrTiO$_3$ and LaAlO$_3$ at room temperature. *Sci. Adv*. **2017,** *3*, e1602312.

(6) Kim, U.; Park, C.; Kim, Y. M.; Shin, J.; Char, K. Conducting interface states at LaInO$_3$/BaSnO$_3$ polar interface controlled by Fermi level. *APL Mater.* **2016,** *4*, 071102.

(7) Kim, Y.; Kim, Y. M.; Shin, J.; Char, K. LaInO$_3$/BaSnO$_3$ polar interface on MgO substrates. *APL Mater.* **2018,** *6*, 096104.

(8) Kim, Y. M.; Markurt, T.; Kim, Y.; Zupancic, M.; Shin, J.; Albrecht, M.; Char, K. Interface polarization model for a 2-dimensional electron gas at the BaSnO$_3$/LaInO$_3$ interface. *Sci. Rep.* **2019,** *9*, 16202.

(9). Shin, J.; Kim, Y. M.; Park, C.; Char, K. Remote Doping of the Two-Dimensional-Electron-Gas State at the LaInO$_3$/BaSnO$_3$ Polar Interface. *Phys. Rev. Appl.* **2020,** *13*, 064066.

(10) Krishnaswamy, K.; Bjaalie, L.; Himmetoglu, B.; Janotti, A.; Gordon, L.; Van de Walle, C. G. BaSnO$_3$ as a channel material in perovskite oxide heterostructures. *Appl. Phys. Lett.* **2016,** *108*, 083501.

(11) Kim, H. J.; Kim, U.; Kim, T. H.; Kim, J.; Kim, H. M.; Jeon, B.-G.; Lee, W.-J.; Mun, H. S.; Hong, K. T.; Yu, J.; Char, K.; Kim, K. H. Physical properties of transparent perovskite oxides (Ba,La)SnO$_3$ with high electrical mobility at room temperature. *Phys. Rev. B* **2012,** *86*, 165205.

(12) Sallis, S.; Scanlon, D. O.; Chae, S. C.; Quackenbush, N. F.; Fischer, D. A.; Woicik, J. C.; Guo, J. H.; Cheong, S. W.; Piper, L. F. J. La-doped BaSnO$_3$—Degenerate perovskite transparent conducting oxide: Evidence from synchrotron x-ray spectroscopy. *Appl. Phys. Lett.* **2013,** *103*, 042105.

(13) Lebens-Higgins, Z.; Scanlon, D. O.; Paik, H.; Sallis, S.; Nie, Y.; Uchida, M.; Quackenbush, N. F.; Wahila, M. J.; Sterbinsky, G. E.; Arena, D. A.; Woicik, J. C.; Schlom, D. G.; Piper, L. F. Direct Observation of Electrostatically Driven Band Gap Renormalization in a Degenerate Perovskite Transparent Conducting Oxide. *Phys. Rev. Lett.* **2016,** *116*, 027602.

(14) Kim, H. J.; Kim, U.; Kim, H. M.; Kim, T. H.; Mun, H. S.; Jeon, B.-G.; Hong, K. T.; Lee, W.-J.; Ju, C.; Kim, K. H.; Char, K. High Mobility in a Stable Transparent Perovskite Oxide. *Appl. Phys. Express* **2012,** *5*, 061102.

(15) Raghavan, S.; Schumann, T.; Kim, H.; Zhang, J. Y.; Cain, T. A.; Stemmer, S. High-mobility BaSnO$_3$ grown by oxide molecular beam epitaxy. *APL Mater.* **2016,** *4*, 016106.

(16) Prakash, A.; Xu, P.; Faghaninia, A.; Shukla, S.; Ager, J. W., 3rd; Lo, C. S.; Jalan, B. Wide bandgap BaSnO$_3$ films with room temperature conductivity exceeding $10^4$ S cm$^{-1}$. *Nat. Commun.* **2017,** *8*, 15167.

(17) Shin, J.; Kim, Y. M.; Kim, Y.; Park, C.; Char, K. High mobility BaSnO$_3$ films and field effect transistors on non-perovskite MgO substrate. *Appl. Phys. Lett.* **2016,** *109*, 262102.

(18) Sanchela, A. V.; Onozato, T.; Feng, B.; Ikuhara, Y.; Ohta, H. Thermopower modulation clarification of the intrinsic effective mass in transparent oxide semiconductor $BaSnO_3$. *Phys. Rev. Mater.* **2017,** *1*, 034603.

(19) Krishnaswamy, K.; Himmetoglu, B.; Kang, Y.; Janotti, A.; Van de Walle, C. G. First-principles analysis of electron transport in $BaSnO_3$. *Phys. Rev. B* **2017,** *95*, 205202.

(20) Popovic, Z. S.; Satpathy, S.; Martin, R. M. Origin of the two-dimensional electron gas carrier density at the $LaAlO_3$ on $SrTiO_3$ interface. *Phys. Rev. Lett.* **2008,** *101*, 256801.

(21) Geller, S. Crystal Structure of Gadolinium Orthoferrite, $GdFeO_3$. *J. Chem. Phys.* **1956,** *24*, 1236-1239.

(22) Kim, U.; Park, C.; Ha, T.; Kim, Y. M.; Kim, N.; Ju, C.; Park, J.; Yu, J.; Kim, J. H.; Char, K. All-perovskite transparent high mobility field effect using epitaxial $BaSnO_3$ and $LaInO_3$. *APL Mater.* **2015,** *3*, 036101.

(23) Ubic, R.; Subodh, G. The prediction of lattice constants in orthorhombic perovskites. *J. Alloys Compd.* **2009,** *488*, 374-379.

(24) Park, H. M.; Lee, H. J.; Park, S. H.; Yoo, H. I. Lanthanum indium oxide from X-ray powder diffraction. *Acta Cryst. C* **2003,** *59*, i131-2.

(25) Chezhina, N. V.; Bodritskaya, É. V.; Zhuk, N. A.; Bannikov, V. V.; Shein, I. R.; Ivanovskiĭ, A. L. Magnetic properties and electronic structure of the $LaGaO_3$ perovskite doped with nickel. *Phys. Solid State* **2008,** *50*, 2121-2126.

(26) Vasylechko, L.; Matkovski, A.; Suchocki, A.; Savytskii, D.; Syvorotka, I. Crystal structure of $LaGaO_3$ and $(La,Gd)GaO_3$ solid solutions. *J. Alloys Compd.* **1999,** *286*, 213-218.

(27) Kim, Y. M.; Kim, Y.; Char, K. The role of coherent epitaxy in forming a two-dimensional electron gas at $LaIn_{1-x}Ga_xO_3/BaSnO_3$ interfaces. *Commun. Mater.* **2021,** *2*, 73.

(28) Christen, H. M.; Jellison, G. E.; Ohkubo, I.; Huang, S.; Reeves, M. E.; Cicerrella, E.; Freeouf, J. L.; Jia, Y.; Schlom, D. G. *Appl. Phys. Lett.* **2006,** *88*, 262906.

(29) Arima, T.; Tokura, Y.; Torrance, J. B. Variation of optical gaps in perovskite-type 3$d$ transition-metal oxides. *Phys. Rev. B* **1993,** *48*, 17006-17009.

(30) Heeg, T.; Wagner, M.; Schubert, J.; Buchal, C.; Boese, M.; Luysberg, M.; Cicerrella, E.; Freeouf, J. L. Rare-earth scandate single- and multi-layer thin films as alternative gate oxides for microelectronic applications. *Microelectron. Eng.* **2005,** *80*, 150-153.

(31) Heeg, T.; Schubert, J.; Buchal, C.; Cicerrella, E.; Freeouf, J. L.; Tian, W.; Jia, Y.; Schlom, D. G. Growth and properties of epitaxial rare-earth scandate thin films. *Appl. Phys. A* **2006,** *83*, 103-106.

(32) Melton, C. A.; Mitas, L. Many-body electronic structure of $LaScO_3$ by real-space quantum Monte Carlo. *Phys. Rev. B* **2020,** *102*, 045103.

(33) Clark, J. B.; Richter, P. W.; du Toit, L. High-Pressure Synthesis of $YScO_3$, $HoScO_3$, $ErScO_3$, and $TmScO_3$, and a Reevaluation of the Lattice Constants of the Rare Earth Scandates. *J. Solid State Chem.* **1978,** *23*, 129-134.

(34) Liferovich, R. P.; Mitchell, R. H. A structural study of ternary lanthanide orthoscandate perovskites. *J. Solid State Chem.* **2004,** *177*, 2188-2197.

(35) Zhao, C.; Witters, T.; Brijs, B.; Bender, H.; Richard, O.; Caymax, M.; Heeg, T.; Schubert, J.; Afanas'ev, V. V.; Stesmans, A.; Schlom, D. G. Ternary rare-earth metal oxide high-$k$ layers on silicon oxide. *Appl. Phys. Lett.* **2005,** *86*, 132903.

(36) Paudel, T. R.; Tsymbal, E. Y. Prediction of a mobile two-dimensional electron gas at the $LaScO_3$/$BaSnO_3$ (001) interface. *Phys. Rev. B* **2017,** *96*, 245423.

(37) Kumar, S.; Kaswan, J.; Satpati, B.; Shukla, A. K.; Gahtori, B.; Pulikkotil, J. J.; Dogra, A. $LaScO_3$/$SrTiO_3$: A conducting polar heterointerface of two 3$d$ band insulating perovskites. *Appl. Phys. Lett.* **2020,** *116*, 051603.

(38) Mun, H.; Kim, U.; Min Kim, H.; Park, C.; Hoon Kim, T.; Joon Kim, H.; Hoon Kim, K.; Char, K. Large effects of dislocations on high mobility of epitaxial perovskite $Ba_{0.96}La_{0.04}SnO_3$ films. *Appl. Phys. Lett.* **2013,** *102*, 252105.

(39) Lenzlinger, M.; Snow, E. H. Fowler-Nordheim Tunneling into Thermally Grown SiO$_2$. *J. Appl. Phys.* **1969,** *40*, 278-283.

(40) Kim, Y. M.; Park, C.; Ha, T.; Kim, U.; Kim, N.; Shin, J.; Kim, Y.; Yu, J.; Kim, J. H.; Char, K. High-k perovskite gate oxide BaHfO$_3$. *APL Mater.* **2017,** *5*, 016104.

(41) Siemons, W.; Koster, G.; Yamamoto, H.; Harrison, W. A.; Lucovsky, G.; Geballe, T. H.; Blank, D. H.; Beasley, M. R. Origin of charge density at LaAlO$_3$ on SrTiO$_3$ heterointerfaces: possibility of intrinsic doping. *Phys. Rev. Lett.* **2007,** *98*, 196802.

(42) Herranz, G.; Basletic, M.; Bibes, M.; Carretero, C.; Tafra, E.; Jacquet, E.; Bouzehouane, K.; Deranlot, C.; Hamzic, A.; Broto, J. M.; Barthelemy, A.; Fert, A. High mobility in LaAlO$_3$/SrTiO$_3$ heterostructures: origin, dimensionality, and perspectives. *Phys. Rev. Lett.* **2007,** *98*, 216803.

(43) Tan, I. H.; Snider, G. L.; Chang, L. D.; Hu, E. L. A self-consistent solution of Schrödinger–Poisson equations using a nonuniform mesh. *J. Appl. Phys.* **1990,** *68*, 4071-4076.

(44) Singh, P.; Brandenburg, B. J.; Sebastian, C. P.; Singh, P.; Singh, S.; Kumar, D.; Parkash, O. Electronic Structure, Electrical and Dielectric Properties of BaSnO$_3$ below 300 K. *Jpn. J. Appl. Phys.* **2008,** *47*, 3540-3545.

(45) Kim, Y. M. Epitaxial perovskite oxide heterostructure using BaSnO$_3$, BaHfO$_3$, and LaIn$_{1-x}$Ga$_x$O$_3$. Ph.D. Thesis, Seoul National University, Seoul, Republic of Korea, **2019.**

(46) Dingle, R.; Störmer, H. L.; Gossard, A. C.; Wiegmann, W. Electron mobilities in modulation-doped semiconductor heterojunction superlattices. *Appl. Phys. Lett.* **1978,** *33*, 665-667.

(47) Störmer, H. L.; Pinczuk, A.; Gossard, A. C.; Wiegmann, W. Influence of an undoped (AlGa)As spacer on mobility enhancement in GaAs-(AlGa)As superlattices. *Appl. Phys. Lett.* **1981,** *38*, 691-693.

(48) Kim, Y.; Kim, S.; Cho, H.; Kim, Y. M.; Ohta, H.; Char, K. Transport properties of LaInO$_3$/BaSnO$_3$ interface analyzed by Poisson-Schrödinger equation. *unpublished.*

(49) McCollam, A.; Wenderich, S.; Kruize, M. K.; Guduru, V. K.; Molegraaf, H. J. A.; Huijben, M.; Koster, G.; Blank, D. H. A.; Rijnders, G.; Brinkman, A.; Hilgenkamp, H.; Zeitler, U.; Maan, J. C. Quantum oscillations and subband properties of the two-dimensional electron gas at the LaAlO$_3$/SrTiO$_3$ interface. *APL Mater.* **2014,** *2*, 022102.

(50) Chang, L. L.; Sakaki, H.; Chang, C. A.; Esaki, L. Shubnikov—de Haas Oscillations in a Semiconductor Superlattice. *Phys. Rev. Lett.* **1977,** *38*, 1489-1493.

(51) Klitzing, K. v.; Dorda, G.; Pepper, M. New Method for High-Accuracy Determination of the Fine-Structure Constant Based on Quantized Hall Resistance. *Phys.Rev. Lett.* **1980,** *45*, 494-497.

(52) Kim, U.; Park, C.; Ha, T.; Kim, R.; Mun, H. S.; Kim, H. M.; Kim, H. J.; Kim, T. H.; Kim, N.; Yu, J.; Kim, K. H.; Kim, J. H.; Char, K. Dopant-site-dependent scattering by dislocations in epitaxial films of perovskite semiconductor BaSnO$_3$. *APL Mater.* **2014,** *2*, 056107.

(53) Zupancic, M.; Aggoune, W.; Markurt, T.; Kim, Y.; Kim, Y. M.; Char, K.; Draxl, C.; Albrecht, M. Role of the interface in controlling the epitaxial relationship between orthorhombic LaInO$_3$ and cubic BaSnO$_3$. *Phys. Rev. Mater.* **2020,** *4*, 123605.

(54) Lau, C.; Kim, Y.; Albright, S.; Char, K.; Ahn, C. H.; Walker, F. J. Structural characterization of the LaInO$_3$/BaSnO$_3$ interface via synchrotron scattering. *APL Mater.* **2019,** *7*, 031108.

(55) Wang, Y.; Sui, R.; Bi, M.; Tang, W.; Ma, S. Strain sensitivity of band structure and electron mobility in perovskite BaSnO$_3$: first-principles calculation. *RSC Adv.* **2019,** *9*, 14072-14077.

(56). Park, C.; Kim, U.; Ju, C. J.; Park, J. S.; Kim, Y. M.; Char, K. High mobility field effect transistor based on BaSnO$_3$ with Al$_2$O$_3$ gate oxide. *Appl. Phys. Lett.* **2014,** *105*, 203503.

(57) Kim, Y. M.; Park, C.; Kim, U.; Ju, C.; Char, K. High-mobility BaSnO$_3$ thin-film transistor with HfO$_2$ gate insulator. *Appl. Phys. Express* **2016,** *9*, 011201.

(58) Park, J.; Paik, H.; Nomoto, K.; Lee, K.; Park, B.-E.; Grisafe, B.; Wang, L.-C.; Salahuddin, S.; Datta, S.; Kim, Y.; Jena, D.; Xing, H. G.; Schlom, D. G. Fully transparent field-effect transistor with high drain current and on-off ratio. *APL Mater.* **2020,** *8*, 011110.

(59) Cheng, J.; Yang, H.; Combs, N. G.; Wu, W.; Kim, H.; Chandrasekar, H.; Wang, C.; Rajan, S.; Stemmer, S.; Lu, W. Electron transport of perovskite oxide $BaSnO_3$ on (110) $DyScO_3$ substrate with channel-recess for ferroelectric field effect transistors. *Appl. Phys. Lett.* **2021,** *118*, 042105.

(60) Yue, J.; Prakash, A.; Robbins, M. C.; Koester, S. J.; Jalan, B. Depletion Mode MOSFET Using La-Doped $BaSnO_3$ as a Channel Material. *ACS Appl. Mater. Interfaces* **2018,** *10*, 21061-21065.